\documentclass[12pt,thmsa]{article}
\usepackage{amssymb,amsfonts,amsthm,dsfont,mathtools,lipsum,etoolbox}
\usepackage{amsmath}
\usepackage{booktabs,caption,fixltx2e}
\usepackage[flushleft]{threeparttable}
\usepackage{multirow}
\usepackage{setspace}
\usepackage{color}
\usepackage{endnotes}
\usepackage{enumerate}
\usepackage{float}
\usepackage[bottom]{footmisc}
\usepackage[pdftex]{graphicx}
\usepackage{epstopdf}
\usepackage{tikz}
\usepackage{indentfirst}
\usepackage{latexsym}
\usepackage{lscape}
\usepackage{parskip}
\usepackage{rotating}
\usepackage{caption}
\usepackage{subfigure}
\usepackage{bm}
\usepackage{verbatim}
\usepackage{mathdots}
\usepackage[authoryear, round,semicolon,sort&compress]{natbib}
\usepackage{soul}
\usepackage[left=1.0in,right=1.0in,top=1.0in,bottom=1.0in]{geometry}
\usepackage{multirow}
\usepackage{mwe}
\usepackage{arydshln}
\usepackage[scaled=0.85]{beramono}
\usepackage{datetime}
\usepackage{titlesec}
\usepackage[colorlinks=true,allcolors=blue]{hyperref}
\allowdisplaybreaks[4]

\newcommand{\zerodisplayskips}{%
	\setlength{\abovedisplayskip}{0.2cm}%
	\setlength{\belowdisplayskip}{0.2cm}%
	\setlength{\abovedisplayshortskip}{0.2cm}%
	\setlength{\belowdisplayshortskip}{0.2cm}}
\appto{\normalsize}{\zerodisplayskips}
\appto{\small}{\zerodisplayskips}
\appto{\footnotesize}{\zerodisplayskips}

\newtheorem{theorem}{Theorem}
\newtheorem{assumption}{Assumption}

\newtheorem{lemma}{Lemma}

\newcommand{\var}{\text{Var}}

\newcommand{\cov}{\text{Cov}}

\makeatletter
\newcommand{\leqnomode}{\tagsleft@true}
\newcommand{\reqnomode}{\tagsleft@false}
\makeatother

\titlespacing\section{0pt}{10pt plus 4pt minus 2pt}{4pt plus 2pt minus 2pt}
\titlespacing\subsection{0pt}{8pt plus 4pt minus 2pt}{4pt plus 2pt minus 2pt}
\titlespacing\subsubsection{0pt}{6pt plus 4pt minus 2pt}{4pt plus 2pt minus 2pt}

\begin{document}

	\sloppy

	\title{Identifying an Earnings Process With Dependent Contemporaneous Income Shocks}
	
	\author{Dan Ben-Moshe\thanks{{\small Department of Economics, Ben-Gurion University of the Negev, PO BOX 653, Beer-Sheva 84105, Israel. Email: dbmster@gmail.com. 
		} }\\
	}
	\date{May 2023}
	\maketitle

	\begin{abstract}

		




		
		

		
		
	This paper proposes a novel approach for identifying coefficients in an earnings dynamics model with arbitrarily dependent contemporaneous income shocks. Traditional methods relying on second moments fail to identify these coefficients, emphasizing the need for nongaussianity assumptions that capture information from higher moments. Our results contribute to the literature on earnings dynamics by allowing models of earnings to have, for example, the permanent income shock of a job change to be linked to the contemporaneous transitory income shock of a relocation bonus.
		

		\vspace{0.5cm}
		
		\noindent \textbf{Keywords:} Earnings dynamics model, statistically dependent contemporaneous income shocks
		
	\end{abstract}

	
	\pagebreak
	
	\section{Introduction}
	
This papers analyzes the  canonical model for the earnings dynamics (ED) process,
\begin{align}
	\left.	\begin{array}{llll}
		y_{it}&= v_{it}+ w_{it} , & \quad  i=1,\ldots,n, \ \ t=1,\ldots,T, \\
		v_{it}& 
		=\eta_{i1} +\ldots +\eta_{it},\\
		w_{it}&= a_q \xi_{it-q}+\ldots+a_1 \xi_{it-1} + \xi_{it},
	\end{array} \right\}
	\label{eq:ed}
\end{align}
where $y_{it}$ is the residual of a regression of log earnings on covariates,
$v_{it}$ is permanent income component modeled as a random walk with permanent shock
$\eta_{it}$, and $w_{it}$ is transitory income component modeled as a moving
average process of order $q$ with transitory shock $\xi_{it}$. 

While previous research has been interested in allowing for unobserved individual-level heterogeneity \citep[e.g.,][]{almuzaraheterogeneity,botoED22,botosaru2018nonparametric,browning_carro_2007,meghir2004income}, most have assumed that contemporaneous permanent and transitory shocks are uncorrelated. In contrast, we focus on the ED model with a more limited degree of unobserved individual-level heterogeneity but allow for correlated contemporaneous shocks. 

Our goal is to show identification of the coefficients in the MA(q) process of the transitory income component, which we show in some cases is not identified  by traditional methods relying on second moments. To achieve this, we introduce nongaussianity assumptions that capture information from higher moments. Our approach can accommodate scenarios where, for example, a job change comes with a relocation bonus that links the permanent shock of a job change with the transitory shock of a relocation bonus.

The techniques used in this paper are similar to those employed in my previous work on identifying measurement error models \citep{ben-moshe2020eiv}. However, the system of equations and dependence assumptions in the measurement error model are different from those in the ED model considered in this paper.\footnote{The measurement error model with a single regressor is $\begin{pmatrix}y_{i1}\\y_{i2}	\end{pmatrix} =\begin{pmatrix} a_1 \\ 1\end{pmatrix} \xi_i + \begin{pmatrix}	\eta_{i1} \\ \eta_{i2}	\end{pmatrix},$	while the closest ED model in this paper is	$				\begin{pmatrix}y_{i1}\\y_{i2}	\end{pmatrix} =\begin{pmatrix} 1& 1 \\ a_1 & 1\end{pmatrix} \begin{pmatrix}\eta_{i1}  \\ \xi_{i1} \end{pmatrix} + \begin{pmatrix}			a_1\xi_{i0} \\ \eta_{i2} + \xi_{i2}		\end{pmatrix}.$} In \cite{ben-moshe2018dmu}, coefficients are assumed to be known and distributions are identified, while in this paper, we identify the coefficients. Once the coefficients in the ED model are identified, the results from my previous work can be used to identify the joint distribution of income shocks.

\section{Identification} \label{se:ED}

This section identifies the coefficients in the ED model \eqref{eq:ed} allowing contemporaneous income shocks to be arbitrarily dependent.
\vspace{0.125cm}
\begin{assumption} \label{asn:ed}
	$\xi_{1-q},\ldots,\xi_{0},(\xi_{1},\eta_{1}),\ldots,(\xi_{T},\eta_{T})$ are mutually independent.
\end{assumption}
\vspace{0.125cm}
\begin{theorem} \label{th:ed}
	Suppose that \eqref{eq:ed} and Assumption \ref{asn:ed} hold. Assume $q+3 \leq T$ and $a_1\notin\{0,1\},a_q\notin\{0,1\}$. If $\xi_1$ is nongaussian then $a_1,\ldots,a_q$ are identified.
\end{theorem}	
\vspace{0.125cm}
The proof involves several steps. First, the system of linear equations is transformed into a functional equation through the log characteristic function. This equation is the sum of log characteristic functions of the income shocks. Then, a sequence of derivatives is taken to eliminate functions. At each step, arguments are chosen reflecting a change in direction that will allow the next derivative to eliminate a function. In these cases, the order in which the derivatives are taken matters, as changing the order can affect the direction and result in the derivative not eliminating a function. However, when the arguments do not change, the order in which the derivatives are taken does not matter. Finally, the resulting system of differential equations can be integrated to show that $\xi_1$ is normal or the coefficients are identified.

Theorem \ref{th:ed} identifies the coefficients in the ED model with dependent contemporaneous shocks. The most common way to identify
coefficients is by second moments but this results in equations that are nonlinear in the parameters. Further in the case that $q=1$,
the following theorem states that second moments do not identify the coefficient $a_1$  no matter how large $T$, so that information from higher moments is required for identification.
\vspace{0.125cm}
\begin{theorem} \label{th:cov}
	Suppose that \eqref{eq:ed} and Assumption \ref{asn:ed} hold. Assume that $q=1$ and $a_1 \neq 0$. 
	If shocks are jointly normal with positive definite covariance matrix then $a_1$ is not identified. 
\end{theorem}	
\vspace{0.125cm}
The system of covariance equations is linear in the unobserved covariances  $\sigma_{\xi_0}^2,\sigma_{\xi_1}^2,\sigma_{\eta_1,\xi_1},\sigma_{\eta_1}^2,\ldots$ An observationally equivalent model is obtained by solving this system of linear equations. 

The following lemma states that under the stronger assumption that $\eta_1$ and $\xi_1$ are independent, identification is possible with $q+1$ time periods, rather than $q+3$  time periods as in Theorem \ref{th:ed}.
\vspace{0.125cm}
\begin{lemma}\label{lma:ind}
	Suppose that \eqref{eq:ed} and Assumption \ref{asn:ed} hold. Assume that $\eta_1$ and $\xi_1$ are independent and  $\xi_1$ is nongaussian. 
	If $q+1 \leq T$ and $a_q \neq 0$ then $a_1,\ldots, a_q$ are identified.
\end{lemma}

\section{Summary}

This paper identifies coefficients in an earnings dynamics process with arbitrarily dependent contemporaneous permanent and transitory shocks.
The coefficients  are in general not identified by second moments and so require a nongaussianity assumption for identification. 

\appendix

\section{Appendix}

The following theorem from \citet{marcinkiewicz1939propriete} states that if the log characteristic function of a random variable is a polynomial in a neighbourhood of the origin then the random variable is normally distributed. 
\vspace{0.1cm}
\begin{theorem} \label{th:Mar} \citep{marcinkiewicz1939propriete}.
	Let $\varphi(.)$ be a polynomial in a neighbourhood of the origin. 	If $\varphi(.)$ is a log characteristic function then $\varphi(.)$ is of degree at most $2$ and corresponds to a normal distribution.
\end{theorem}
\vspace{0.1cm}
\hspace{-0.01cm} \textbf{Proof of Theorem \ref{th:ed}:} \label{ap:ed}
Let $(\widetilde a_1,\ldots,\widetilde a_q,\widetilde \xi_{1-q},\ldots,\widetilde \xi_T,\widetilde \eta_1,\ldots,\widetilde \eta_T)$ be observationally equivalent to $(a_1,\ldots,a_q,\xi_{1-q},\ldots,\xi_T,\eta_1,\ldots,\eta_T)$.
The log characteristic function of \eqref{eq:ed} is,
\begin{align*}
	&\varphi_{\bm Y}(\bm s)
	=\varphi_{\xi_{1-q}}(a_qs_1)+ \ldots+\varphi_{\xi_0}(\sum_{j=1}^q a_js_{j})
	+ \varphi_{\eta_{1},\xi_{1}}(\sum_{t=1}^{T} s_t, s_1+\sum_{j=1}^q a_js_{j+1})\\
	&+  \varphi_{\eta_{2},\xi_{2}}(\sum_{t=2}^{T} s_t, s_2+\sum_{j=1}^q a_js_{j+2}) 
	+\ldots 
	+ \varphi_{\eta_{T-1},\xi_{T-1}}(s_{T-1}+s_T,s_{T-1}+a_1 s_T)
	+ \varphi_{\eta_{T}+\xi_{T}}(s_T)\\ 
	&=\varphi_{\widetilde\xi_{1-q}}(\widetilde a_qs_1)+ \ldots+ \varphi_{\widetilde\xi_0}(\sum_{j=1}^q \widetilde a_js_{j})
	+ \varphi_{\widetilde\eta_{1},\widetilde\xi_{1}}(\sum_{t=1}^{T} s_t, s_1+\sum_{j=1}^q \widetilde a_js_{j+1})\\
	&+  \varphi_{\widetilde\eta_{2},\widetilde\xi_{2}}(\sum_{t=2}^{T} s_t, s_2+\sum_{j=1}^q \widetilde a_js_{j+2}) 
	+\ldots 
	+ \varphi_{\widetilde\eta_{T-1},\widetilde\xi_{T-1}}(s_{T-1}+s_T,s_{T-1}+a_1 s_T)
	+  \varphi_{\widetilde \eta_{T}+\widetilde\xi_{T}}(s_T),
\end{align*}
For argument $u \in \mathbb{R}$, consider the following derivatives,
{\begin{align*} 
		&\frac{d}{d s_2} \frac{d}{d s_2} \frac{d}{d s_1} \varphi_{\bm Y}(\bm s)\Big|_{(s_1,\ldots,s_{j+1},\ldots,s_{q+3})=(\widetilde a_j,\bm 0-1,\bm 0, 1-\widetilde a_j)u}\\
		&=\Big(\frac{\partial^3}{\partial x_1^3} + (1+2a_1)\frac{\partial^3}{\partial x_1^2\partial x_2}+ (a_1^2+2a_1)\frac{\partial^3}{\partial x_1\partial x_2^2}+a_1^2\frac{\partial^3}{\partial x_2^3}\Big)  \varphi_{\eta_{1},\xi_{1}}(\bm x) \Big|_{\bm x=(0, \widetilde a_j -a_j)u}\\
		&=\Big(\frac{\partial^3}{\partial x_1^3} + (1+2\widetilde a_1)\frac{\partial^3}{\partial x_1^2\partial x_2}+ (\widetilde a_1^2+2\widetilde a_1)\frac{\partial^3}{\partial x_1\partial x_2^2}+\widetilde a_1^2\frac{\partial^3}{\partial x_2^3}\Big)   \varphi_{\widetilde \eta_{1},\widetilde \xi_{1}}(\bm x) \Big|_{\bm x=(0,0)},\\
		&\frac{d}{d s_{q+2}} \frac{d}{d s_{1}} \frac{d}{d s_1} \varphi_{\bm Y}(\bm s)\Big|_{(s_1,\ldots,s_{j+1},\ldots,s_{q+3})=(\widetilde a_j,\bm 0-1,\bm 0, 1-\widetilde a_j)u}\\
		&=\Big(\frac{\partial^3}{\partial x_1^3} + 2\frac{\partial^3}{\partial x_1^2\partial x_2}+\frac{\partial^3}{\partial x_1\partial x_2^2}\Big)  \varphi_{\eta_{1},\xi_{1}}(\bm x) \Big|_{\bm x=(0, \widetilde a_j -a_j)u}\\
		&=\Big(\frac{\partial^3}{\partial x_1^3} + 2\frac{\partial^3}{\partial x_1^2\partial x_2}+\frac{\partial^3}{\partial x_1\partial x_2^2}\Big)  \varphi_{\widetilde \eta_{1},\widetilde \xi_{1}}(\bm x) \Big|_{\bm x=(0,0)},\\ 
		&\frac{d}{d s_{q+2}} \frac{d}{d s_{q+2}} \frac{d}{d s_1} \varphi_{\bm Y}(\bm s)\Big|_{(s_1,\ldots,s_{j+1},\ldots,s_{q+3})=(\widetilde a_j,\bm 0-1,\bm 0, 1-\widetilde a_j)u}\\
		&=\Big(\frac{\partial^3}{\partial x_1^3} + \frac{\partial^3}{\partial x_1^2\partial x_2}\Big)  \varphi_{\eta_{1},\xi_{1}}(\bm x) \Big|_{\bm x=(0, \widetilde a_j -a_j)u} 
		=\Big(\frac{\partial^3}{\partial x_1^3} + \frac{\partial^3}{\partial x_1^2\partial x_2}\Big)   \varphi_{\widetilde \eta_{1},\widetilde \xi_{1}}(\bm x) \Big|_{\bm x=(0,0)},\\
		&\frac{d}{d w_{q+3}}\frac{d}{d v_{q+3}}\frac{d}{d s_2} \varphi_{\bm Y}(\bm s) \Big|_{\bm s=(v_{q+2},\frac{v_2}{1-\widetilde a_{q}}-\widetilde a_{q}v_{q+3},v_3,\ldots,v_{q+1},-\frac{v_2}{1-\widetilde a_{q}}+v_{q+3},v_1+(\widetilde a_{q}-1)v_{q+3})} \\
		&\hspace{1.5cm} \Big|_{\bm v=(w_1,\frac{w_{q+3}(1-\widetilde a_{q})}{1-a_{q}}, w_{3},\ldots,w_{q+1}, w_{q+2},\frac{w_2-w_{q+3}}{a_{q}-\widetilde a_{q}})}  
		\Big|_{(w_1,w_{2},\ldots,w_{q+2},w_{q+3})=(-\widetilde a_q,\frac{a_q-\widetilde a_q}{\widetilde a_q},\bm 0,\widetilde a_q,0)u} \\
		&=\frac{a_1^2 \widetilde a_qa_q(\widetilde a_q-1)}{(1-a_q)(a_q-\widetilde a_q)}  \Big(\frac{\partial ^3}{\partial x_1\partial x_2^2} 
		+ a_1\frac{\partial^3 }{\partial x_2^3}\Big)\varphi_{\eta_{1},\xi_{1}}(\bm x)\Big|_{\bm x=(0,{\widetilde a_1- a_1})u}  \\
		&=\frac{\widetilde a_1 ^2\widetilde a_qa_q(\widetilde a_q-1)}{(1-a_q)(a_q-\widetilde a_q)}\Big(\frac{\partial^3}{\partial x_1\partial x_2^2}
		+ \widetilde a_1\frac{\partial^3}{\partial x_2^3}\Big) \varphi_{\widetilde \eta_{1},\widetilde \xi_{1}}(\bm x)\Big|_{\bm x=(0,0)}  , \\
		&\frac{d}{d w_{q+3}}\frac{d}{d v_{q+3}}\frac{d}{d s_2} \varphi_{\bm Y}(\bm s) \Big|_{\bm s=(v_{q+2},\frac{v_2}{1-\widetilde a_{q}}-\widetilde a_{q}v_{q+3},v_3,\ldots,v_{q+1},-\frac{v_2}{1-\widetilde a_{q}}+v_{q+3},v_1+(\widetilde a_{q}-1)v_{q+3})} \\
		&\hspace{1.5cm} \Big|_{\bm v=(w_1,\frac{w_{q+3}(1-\widetilde a_{q})}{1-a_{q}}, w_{3},\ldots,w_{q+1}, w_{q+2},\frac{w_2-w_{q+3}}{a_{q}-\widetilde a_{q}})}  
		\Big|_{(w_1,\ldots,w_{j+1},\ldots,w_{q+2},w_{q+3})=(1-\widetilde a_j,\bm 0,-1,\bm 0,\widetilde a_j,0)u} \\
		&=\frac{a_1^2 \widetilde a_qa_q(\widetilde a_q-1)}{(1-a_q)(a_q-\widetilde a_q)}  \Big(\frac{\partial ^3}{\partial x_1\partial x_2^2} 
		+ a_1\frac{\partial^3 }{\partial x_2^3}\Big)\varphi_{\eta_{1},\xi_{1}}(\bm x)\Big|_{\bm x=(0,{\widetilde a_j- a_j})u}  \qquad j>1\\
		&=\frac{\widetilde a_1 ^2\widetilde a_qa_q(\widetilde a_q-1)}{(1-a_q)(a_q-\widetilde a_q)}\Big(\frac{\partial^3}{\partial x_1\partial x_2^2}
		+ \widetilde a_1\frac{\partial^3}{\partial x_2^3}\Big)  \varphi_{\widetilde \eta_{1},\widetilde \xi_{1}}(\bm x)\Big|_{\bm x=(0,0)}  . 
	\end{align*}
}Noting that the derivatives evaluated at $(0,0)$ are constants, we obtain the linear system,
\begin{align*}
	\begin{pmatrix}
		1& 1+2a_1& a_1^2+2a_1& a_1^2\\
		1&2 &1 &0 \\
		1& 1& 0& 0 \\
		0&0 & 1&a_1
	\end{pmatrix}	
	\begin{pmatrix}
		\frac{\partial^3 }{\partial x_1^3} \varphi_{\eta_{1},\xi_{1}}(\bm x)\Big|_{\bm x=(0,{\widetilde a_j- a_j})u} \\
		\frac{\partial^3 }{\partial x_1^2\partial x_2} \varphi_{\eta_{1},\xi_{1}}(\bm x)\Big|_{\bm x=(0,{\widetilde a_j- a_j})u} \\
		\frac{\partial^3 }{\partial x_1\partial x_2^2} \varphi_{\eta_{1},\xi_{1}}(\bm x)\Big|_{\bm x=(0,{\widetilde a_j- a_j})u} \\
		\frac{\partial^3 }{\partial x_2^3} \varphi_{\eta_{1},\xi_{1}}(\bm x)\Big|_{\bm x=(0,{\widetilde a_j- a_j})u} 
	\end{pmatrix}&= \bm C.
\end{align*}
The matrix on the left is invertible as long as $a_1\notin\{0,1\}$. So, 
\begin{align*}
	\frac{\partial^3 }{\partial x_2^3} \varphi_{\eta_{1},\xi_{1}}(\bm x)\Big|_{\bm x=(0,{\widetilde a_j- a_j})u} &=
	\frac{\partial^3 }{\partial u^3} 	\varphi_{(\widetilde a_j- a_j)\xi_1}(u) =c. 
\end{align*}
Hence, $\varphi_{(\widetilde a_j- a_j)\xi_1}(u)$ is a polynomial.
By \cite{marcinkiewicz1939propriete}, $(\widetilde a_j- a_j)\xi_1$ is normal. By assumption $\xi_1$ is nongaussian, so $\widetilde a_j=a_j$ and $a_j$ is identified.    
\hfill$\square$ 

\vspace{0.05cm}

\textbf{Proof of Theorem \ref{th:cov}:} \label{ap:cov}
Let $(a_1,\xi_0,\ldots,\xi_T,\eta_1,\ldots,\eta_T)$ be observationally equivalent to $(\widetilde a_1,\widetilde \xi_0,,\ldots,\widetilde \xi_T,\widetilde \eta_1,\ldots,\widetilde \eta_T)$.
Since the shocks are jointly normal (with mean zero) all the information is in the system of second moments:
\begin{align*}
	\var(y_1)&= a_1^2\sigma_{\xi_{0}}^2+\sigma_{\eta_1}^2+\sigma_{\xi_{1}}^2+2\sigma_{\eta_1\xi_1}
	=\widetilde a_1^2\sigma_{\widetilde \xi_{0}}^2+ \sigma_{\widetilde \eta_1}^2+ \sigma_{\widetilde\xi_{1}}^2+2\sigma_{\widetilde\eta_1\widetilde\xi_1} , \\
	\cov(y_{1},y_{2})&=\sigma_{\eta_1}^2 +a_1\sigma_{\xi_{1}}^2+(1+a_1)\sigma_{\eta_{1}\xi_{1}}
	=\sigma_{\widetilde\eta_1}^2 +\widetilde a_1\sigma_{\widetilde \xi_{1}}^2+(1+\widetilde a_1)\sigma_{\widetilde \eta_{1}\widetilde \xi_{1}},\\
	\cov(y_{1},y_{1+j})&=\sigma_{\eta_1}^2+\sigma_{\eta_{1}\xi_{1}} 
	=\sigma_{\widetilde \eta_1}^2+\sigma_{\widetilde \eta_{1}\widetilde \xi_{1}}  ,&&   j\geq 2,\\ 
	\var(y_t)&= \sigma_{\eta_1}^2+\ldots + \sigma_{\eta_t}^2  +a_1^2\sigma_{\xi_{t-1}}^2+ \sigma_{\xi_t}^2 +2a_1\sigma_{\eta_{t-1}\xi_{t-1}}+2\sigma_{\eta_t\xi_t}\\
	&= \sigma_{\widetilde \eta_1}^2+\ldots + \sigma_{\widetilde \eta_t}^2  +\widetilde a_1^2\sigma_{\widetilde \xi_{t-1}}^2+ \sigma_{\widetilde \xi_t}^2 +2\widetilde a_1\sigma_{\widetilde \eta_{t-1}\widetilde \xi_{t-1}}+2\sigma_{\widetilde \eta_t\widetilde \xi_t}, && t> 1,\\ 
	\cov(y_{t},y_{t+1})&=\sigma_{\eta_1}^2+\ldots + \sigma_{\eta_{t}}^2  +a_1\sigma_{\xi_{t}}^2+a_1\sigma_{\eta_{t-1}\xi_{t-1}}+(1+a_1)\sigma_{\eta_{t}\xi_{t}}\\
	&=\sigma_{\widetilde \eta_1}^2+\ldots + \sigma_{\widetilde \eta_{t}}^2  +\widetilde a_1\sigma_{\widetilde \xi_{t}}^2+\widetilde a_1\sigma_{\widetilde \eta_{t-1}\widetilde \xi_{t-1}}+(1+\widetilde a_1)\sigma_{\widetilde \eta_{t}\widetilde \xi_{t}}, \\
	\cov(y_{t},y_{t+j})&=\sigma_{\eta_1}^2+\ldots + \sigma_{\eta_{t}}^2  +a_1\sigma_{\eta_{t-1}\xi_{t-1}}+\sigma_{\eta_{t}\xi_{t}}\\
	&=\sigma_{\widetilde \eta_1}^2+\ldots + \sigma_{\widetilde \eta_{t}}^2  +\widetilde a_1\sigma_{\widetilde \eta_{t-1}\widetilde \xi_{t-1}}+\sigma_{\widetilde \eta_{t}\widetilde \xi_{t}},  && j\geq 2.
\end{align*}  
The observationally equivalent model in matrix notation is,
\begin{align*}
	\begin{pmatrix}
		\var(y_1) \\ \cov(y_{1},y_{2}) \\ 	\cov(y_{1},y_{1+j}) \\ \vdots \\	\var(y_t) \\ \cov(y_{t},y_{t+1}) \\ \cov(y_{t},y_{t+j}) \\  \vdots
	\end{pmatrix} &= \begin{pmatrix}	\widetilde a_1^2 & 1 & 2 & 1 &\ldots & 0 & 0 & 0 & \ldots \\
		0 & 1 & 1 +\widetilde a_1& \widetilde a_1 &\ldots  & 0 & 0 & 0 & \ldots \\
		0 & 1 & 1 & 0 &\ldots & 0 & 0 & 0 & \ldots \\
		\vdots & \vdots & \vdots & \vdots & \vdots & \vdots & \vdots & \vdots & \vdots  \\
		0 & 1 & 0& 0&\ldots  & 1 & 2 & 1 & \ldots \\
		0 & 1 & 0& 0&\ldots  & 1 & 1 + \widetilde a_1 & \widetilde a_1 & \ldots \\
		0 & 1 & 0& 0&\ldots  & 1 & 1 & 0 & \ldots \\
		\vdots & \vdots & \vdots & \vdots & \vdots & \vdots & \vdots & \vdots & \vdots  
	\end{pmatrix}   \begin{pmatrix}	\sigma_{\widetilde \xi_{0}}^2 \\ \sigma_{\widetilde \eta_{1}}^2 \\ \sigma_{\widetilde\eta_1\widetilde\xi_1} \\ \sigma_{\widetilde \xi_{1}}^2  \\ \vdots \\ 	\sigma_{\widetilde \eta_{t}}^2 \\ \sigma_{\widetilde\eta_t\widetilde\xi_t} \\ \sigma_{\widetilde \xi_{t}}^2 \\  \vdots \end{pmatrix}  .
\end{align*}  
Given any $\widetilde a_1 \neq 0$, this is a system of linear equations with full row rank, so there are always covariances $\sigma_{\widetilde \xi_{0}}^2,\sigma_{\widetilde \eta_{1}}^2,\sigma_{\widetilde \eta_1\widetilde \xi_1},\sigma_{\widetilde \xi_{1}}^2,\ldots$ that solve these equations. By continuity of solutions for linear systems, we can always find $\widetilde a_1$ close enough to $a_1$ so that the observationally equivalent shocks have positive definite covariance matrix. Thus the model is not identified. \hfill$\square$ 

\vspace{0.05cm}

\textbf{Proof of Lemma \ref{lma:ind}:} \label{ap:ind}
The log characteristic function of \eqref{eq:ed} is,
\begin{align*}
	&	\varphi_{\bm Y}(\bm s)
	=\varphi_{\xi_{1-q}}(a_qs_1)+ \ldots+\varphi_{\xi_0}(\sum_{j=1}^q a_js_{j})
	+ \varphi_{\xi_1}(s_1+\sum_{j=1}^q a_js_{j+1}) 	+ \varphi_{\eta_1}(\sum_{t=1}^{T} s_t)\\
	&+  \varphi_{\eta_2,\xi_2}(\sum_{t=2}^{T} s_t, s_2+\sum_{j=1}^q a_js_{j+2})  
	+\ldots 
	+ \varphi_{\eta_{T-1},\xi_{T-1}}(s_{T-1}+s_T,s_{T-1}+a_1 s_T)
	+ \varphi_{\eta_T+\xi_T}(s_T).
\end{align*}
Consider the following derivatives,
\begin{align*}
	\frac{\partial^2}{\partial s_1 \partial s_{q+1}}\varphi_{\bm Y}(\bm s)&=
	a_q	\frac{\partial^2}{\partial u^2}
	\varphi_{\xi_1}(u)\Big|_{u=s_1+\sum_{j=1}^q a_js_{j+1}}+\frac{\partial^2}{\partial u^2}\varphi_{\eta_1}(u)\Big|_{u=\sum_{t=1}^{T} s_t}.
\end{align*}
Then for argument $u \in \mathbb{R}$,
\begin{align*}
	\cov(y_1,y_{q+1})+\frac{\partial^2}{\partial s_1\partial s_{q+1}}\varphi_{\bm Y}(\bm s)\Big|_{\bm s=(\widetilde a_j,\bm 0, -1, \bm 0,1-\widetilde a_j)u}&=a_q\Big(\var(\xi_j)+	\frac{\partial^2}{\partial u^2}\varphi_{(\widetilde a_j - a_j)\xi_1}(u )\Big),
\end{align*}
where $	\cov(y_1,y_{q+1})=-\frac{\partial}{\partial s_1 \partial s_{q+1}}\varphi_{\bm Y}(\bm s)\Big|_{\bm s=\bm 0}$.
Set the equation to be zero,
\begin{align*}
	a_q\Big(\var(\xi_1) +	\frac{\partial^2}{\partial u^2}\varphi_{(\widetilde a_j - a_j)\xi_1}(u )\Big)&=0, && j=1,\ldots,q.
\end{align*}
By assumption $a_q \neq 0$ and $\xi_1$ is nongaussian. Hence, the equation is zero if and only if $\widetilde{a}_j= a_j$. So $a_1,\ldots,a_q$ are identified.
\hfill$\square$  


\end{document}